\documentclass[aps,prl,reprint,groupedaddress]{revtex4-1}
\usepackage{graphicx} 
\usepackage{dcolumn} 
\usepackage{bm}
\usepackage{hyperref}

\begin{document}

\title{ Preequilibrium cluster emission in massive transfer reactions near Coulomb barrier energy }
\author{Zhao-Qing Feng}
\email{Corresponding author: fengzhq@scut.edu.cn}

\affiliation{School of Physics and Optoelectronics, South China University of Technology, Guangzhou 510640, China }

\date{\today}

\begin{abstract}

Within the framework of the dinuclear system model, the preequilibrium emission of neutron, proton, deuteron, triton, $^{3}$He, $\alpha$, $^{6}$Li, $^{7}$Li, $^{8}$Be and $^{9}$Be in the transfer reactions of $^{12}$C + $^{209}$Bi, $^{40,48}$Ca+$^{238}$U, $^{238}$U+$^{238}$U and $^{238}$U+$^{248}$Cm has been systematically investigated. The production rate, kinetic energy spectra and emission angular distribution are calculated. It is found that the preequilibrium emission mechanism is associated with the reaction system and beam energy. The preequilibrium cross sections of proton, deuteron, triton and alpha are comparable in magnitude. The reaction with $^{40}$Ca is favorable for the cluster emission in comparison with $^{48}$Ca on $^{238}$U at the near barrier energy. A broad angular distribution of the preequilibrium cluster is found in the heavy systems $^{238}$U+$^{238}$U and $^{238}$U+$^{248}$Cm.

\begin{description}
\item[PACS number(s)]
25.70.Hi, 25.70.Lm, 24.60.-k      
\end{description}
\end{abstract}

\maketitle

\section{I. Introduction}

The preequilibrium cluster emission in transfer reactions is of significance for the investigation of the correlation of spatial configuration of nucleons, nuclear structure and reaction dynamics. The cluster structure exists in a nucleus, i.e., $^{6}$Li being composed of $\alpha$ and d, $^{8}$Be being the two $\alpha$, three $\alpha$ in $^{12}$C, surface cluster in heavy nucleus etc. The cluster is considered to be formed by the overlap of the singe-particle wave function. In nuclear reactions, it has been known that the preequilibrium cluster formation is different with the one from the decay of the compound nucleus formed in fusion reactions. The cluster emission provides important information on the single-particle or multiparticle correlation of nuclear states, which has been widely used as powerful nuclear spectroscopic tool \cite{Ho03}. On the other hand, the characteristics of the multi-nucleon transfer reaction (MNT) and deep inelastic heavy-ion collisions is related to the preequilibrium cluster emission, which has been attempted to create the neutron-rich heavy nuclei, in particular around the neutron shell closure. There are a number of experiments for measuring the kinetic energy spectra, double differential cross section, angular distribution etc \cite{Ji80,Le82,Di97,Co00,He06,Bu08}. The deep investigation of the preequilibrium cluster emission in transfer reactions is helpful for exploring the cluster structure of the stable or unstable nuclide, the cluster formation mechanism in nuclear reaction, the synthesis of superheavy nucleus or new isotope, the MNT mechanism etc \cite{At71,At73,Hi77,Oe03,Ca06}.

The cluster structure in a nucleus is usually studied by the direct reactions, e.g., the pick-up or knock-out reaction, the breakup reaction etc. The preequilibrium cluster in the massive transfer reaction is also a direct process, in which the cluster is created before the formation of compound nucleus. Both the cluster structure and reaction dynamics influence the preequilibrium cluster production. There are many nuclear models for describing the cluster structure and nucleon condensation in a nucleus, in which the cluster preformation probability in the spatial coordinate can be estimated \cite{Ia82,Ja83,Pi84}. However, the cluster emission in nuclear reaction is very complicated, which is dependent on the space-time evolution, beam energy, cluster structure etc. A few of reaction models or phenomenological formula are proposed for the preequilibrium particle emission, e.g, the exciton model \cite{Bl75,Ga77,Fo10}, Langevin equations \cite{Ji07}, quantum molecular dynamics model \cite{Ch21}. Sophisticated model is still needed for precisely describing the preequilibrium cluster emission. Both the cluster configuration of collision system and reaction dynamics influence the cluster production, i.e., the kinetic energy spectra, angular distribution.

In this work, the preequilibrium cluster emission in the transfer reactions is to be systematically investigated within the framework of dinuclear system (DNS) model. The article is organized as follows. In Section II we give a brief description of the DNS model for describing the preequilibrium cluster production. In Section III, the production cross sections, kinetic energy spectra and angular distribution of the preequilibrium cluster are analyzed and discussed. Summary and perspective on the cluster emission in the transfer reactions are shown in Section IV.

\section{II. Brief description of the model}

The DNS concept was assumed that the colliding system is formed at the touching configuration in nuclear collisions and proposed by Volkov at Dubna for describing the deep inelastic heavy-ion collisions \cite{Vo20}. The typical sticking time is several zeptoseconds. The nucleon exchange and energy dissipation take place once the DNS is formed. The nucleon transfer between the binary fragments is governed by the single-particle Hamiltonian. The DNS model has been used for describing the massive fusion-evaporation mechanism and multi-nucleon transfer reactions \cite{Fe06,Fe07,Fe09}. In this work, the preequilibrium cluster production is to be investigated with the model. The cross sections of the preequilibrium clusters ($\nu=n, p, d, t$, $^{3}$He, $\alpha$, $^{6,7}$Li and $^{8,9}$Be) are estimated as follows
\begin{eqnarray}
	\sigma_{\nu}(E_{k},\theta,t)  = && \sum^{J_{max}}_{J=0,Z_{1},N_{1}} \sigma_{cap}(E_{c.m.},J) \int f(B)   \nonumber\\
	&&  \times  P(Z_{1},N_{1},E_{1}(E_{c.m.},J), t, B)                   \nonumber \\
	&&  \times P_{\nu}(Z_{\nu},N_{\nu},E_{k}) dB.
\end{eqnarray}
Here, $E_{1}$ is the excitation energy for the fragment with $(Z_{1},N_{1})$, respectively, which is associated with the center-of-mass energy $E_{c.m.}$ and incident angular momentum $J$. The maximal angular momentum $J_{max}$ is taken to be the grazing collision of two colliding nuclei. The kinetic energy of cluster is sampled by the Monte Carlo approach within the excitation energy $E_{1}$ . The capture cross section is given by $\sigma_{cap} =\pi\hbar^{2}(2J+1) T(E_{c.m.},J)/(2\mu E_{c.m.})$ with $T(E_{c.m.},J)=\int f(B)T(E_{c.m.},J,B) dB$. The transmission probability $T(E_{c.m.},J,B)$ is calculated by the well known Hill-Wheeler formula for the light and medium systems. For the heavy systems, for example, $^{238}$U+$^{238}$U etc, the classical trajectory approach with a barrier distribution by $T(E_{c.m.},J,B)=0$ and 1 for $E_{c.m.} < B+J(J+1)\hbar^{2}/(2\mu R^{2}_{C})$ and $E_{c.m.} > B+J(J+1)\hbar^{2}/(2\mu R^{2}_{C})$, respectively. The $\mu$ and $R_{C}$ denote the reduced mass and Coulomb radius by $\mu = m_{n} A_{p} A_{t} /(A_{p} + A_{t})$  with $m_{n}$, $A_{p}$ and $A_{t}$ being the nucleon mass and numbers of projectile and target nuclides, respectively. The distribution function is taken as the Gaussian form $f(B) = \frac{1}{N} exp[-((B-B_{m})/\Delta)^{2}]$, with the normalization constant satisfying the unity relation $\int f(B)dB=1$. The quantities $B_{m}$ and $\Delta$ are evaluated by $B_{m}=(B_{C} + B_{S})/2$ and $\Delta = (B_{C} - B_{S})/2$, respectively. The $B_{C}$ and $B_{S}$ are the Coulomb barrier at waist-to-waist orientation and the minimum barrier by varying the quadrupole deformation of the colliding partners.

The nucleon transfer is described by solving a set of microscopically derived master equations by distinguishing protons and neutrons \cite{Fe06,Fe07}. The time evolution of the distribution probability $P(Z_{1},N_{1},E_{1},t)$ for the DNS fragment 1 with proton number $Z_{1}$ and neutron number $N_{1}$ and excitation energy $E_{1}$ is governed by the master equations as follows,
\begin{eqnarray}
&&  \frac{d P(Z_{1},N_{1},E_{1}, t)}{dt}   \nonumber\\
   &&  = \sum_{Z^{'}_{1}}W_{Z_{1},N_{1};Z^{'}_{1},N_{1}}(t)  \nonumber\\              
	&& \times[d_{Z_{1},N_{1}}P(Z^{'}_{1},N_{1},E^{'}_{1},t)     \nonumber\\
	&& - d_{Z^{'}_{1},N_{1}}P(Z_{1},N_{1},E_{1},t)]       \nonumber\\
	&& + \sum_{N^{'}_{1}}W_{Z_{1},N_{1};Z_{1},N^{'}_{1}}(t)[d_{Z_{1},N_{1}}P(Z_{1},N^{'}_{1},E^{'}_{1},t) \nonumber\\
	&& -d_{Z_{1},N^{'}_{1}}P(Z_{1},N_{1},E_{1}, t)]
\end{eqnarray}
Here the $W_{Z_{1},N_{1};Z_{1}^{\prime},N_{1}}$ ($W_{Z_{1},N_{1};Z_{1},N_{1}^{\prime}}$) is the mean transition probability from the channel $(Z_{1},N_{1},E_{1})$ to $(Z_{1}^{\prime},N_{1},E_{1}^{\prime})$ (or $(Z_{1},N_{1},E_{1})$ to $(Z_{1},N_{1}^{\prime},E_{1}^{\prime})$), and $d_{Z_{1},N_{1}}$
denotes the microscopic dimension corresponding to the macroscopic state $(Z_{1},N_{1},E_{1})$. The cascade nucleon transfer is considered in the process with the relation of $Z_{1}^{\prime}=Z_{1}\pm 1$ and $N_{1}^{\prime }=N_{1}\pm 1$. It is noticed that the quasifission of DNS and the fission of heavy fragments are neglected in the dissipation process. Different with the fusion-evaporation reactions \cite{Fe07}, the interaction time is short for the MNT reactions at the level of several zeptoseconds. The interaction time $\tau_{int}$ is obtained from the deflection function method \cite{Wo78}, which depends on the relative angular momentum and colliding system. On the other hand, the interaction potential is flat at the touching distance and no potential pocket exists in the heavy systems. So the quasifission barrier does not appear. We assume the quasifission and fission do not take place before the dissipation equilibrium. The initial probabilities of projectile and target nuclei are set to be $P(Z_{proj},N_{proj},E_{1}=0, t=0)=0.5$ and $P(Z_{targ},N_{targ},E_{1}=0, t=0)=0.5$. The unitary condition is satisfied during the nucleon transfer process $\sum_{Z_{1},N_{1}} P(Z_{1}, N_{1}, E_{1}, t)=1$. The motion of nucleons in the interacting potential is governed by the single-particle Hamiltonian \cite{Fe06}. The excited DNS opens a valence space in which the valence nucleons have a symmetrical distribution around the Fermi surface. Only the particles at the states within the valence space are actively at excitation and transfer. The averages on these quantities are performed in the valence space as follows.
\begin{eqnarray}
\Delta \varepsilon_K = \sqrt{\frac{4\varepsilon^*_K}{g_K}},\quad
\varepsilon^*_K =\varepsilon^*\frac{A_K}{A}, \quad
g_K = A_K /12,
\end{eqnarray}
where the $\varepsilon^*$ is the local excitation energy of the DNS. The microscopic dimension for the fragment ($Z_{K},N_{K}$) is evaluated by the valence states $N_K$ = $g_K\Delta\varepsilon_K$ and the valence nucleons $m_K$ = $N_K/2$ ($K=1,2$) as
\begin{eqnarray}
 d(m_1, m_2) = {N_1 \choose m_1} {N_2 \choose m_2}.
\end{eqnarray}

The transition probability is related to the local excitation energy and nucleon transfer, which is microscopically derived from the interaction potential in valence space as
\begin{eqnarray}
W_{Z_{1},N_{1};Z_{1}^{\prime},N_{1}} && = \frac{\tau_{mem}(Z_{1},N_{1},E_{1};Z_{1}^{\prime},N_{1},E_{1}^{\prime})}{d_{Z_{1},N_{1}} d_{Z_{1}^{\prime},N_{1}}\hbar^{2}}    \nonumber \\
&& \times \sum_{ii^{\prime}}|\langle  Z_{1}^{\prime},N_{1},E_{1}^{\prime},i^{\prime}|V|Z_{1},N_{1},E_{1},i \rangle|^{2}.
\end{eqnarray}
 The memory time is calculated by
\begin{equation}
\tau_{mem}(Z_1,N_1,E_1; Z'_1,N_1, E'_1)=  \left[\frac{2\pi \hbar^2} {\sum _{KK'} <V_{KK} V^*_{KK'}>}\right]^{1/2},
\end{equation}
\begin{eqnarray}
<V_{KK} V^*_{KK'}> && = \frac{1}{4} U^2_{KK'}g_K g_K' \Delta_{KK'} \Delta \varepsilon_K \Delta \varepsilon_K^{\prime}
 \nonumber \\
&&  \times \left[ \Delta^2_{KK'}+ \frac{1}{6} ((\Delta \varepsilon_K)^2  + (\Delta \varepsilon_K^{\prime})^2) \right]^{-1/2}
\end{eqnarray}
The interaction matrix element is given by
\begin{eqnarray}
\sum _{ii'} |V_{ii'}|^2  &&=  [ \omega_{11}(Z_1,N_1,E_1,E'_1)                               \nonumber \\
&&  + \omega_{22}(Z_1,N_1,E_1,E'_1)] \delta_{Z_1,N_1,E_1;Z_1,N_1,E'_1}   \nonumber \\
&& + \omega_{12}(Z_1,N_1,E_1,E'_1)\delta_{Z'_1,N_1,E_1;Z_1-1,N_1,E'_1}   \nonumber \\
&& + \omega_{21}(Z_1,N_1,E_1,E'_1) \delta_{Z'_1,N_1,E_1;Z_1+1,N_1,E'_1}
\end{eqnarray}
with the relation of
\begin{eqnarray}
\omega_{KK'} (Z_1,N_1,E_1,E'_1)=d_{Z_1,N_1} <V_{KK'},V^*_{KK'}>
\end{eqnarray}
The similar process for neutron transfer takes place.

In the relaxation process of the relative motion, the DNS will be excited by the dissipation of the relative kinetic energy. The local excitation energy is determined by the dissipation energy from the relative motion and the potential energy surface of the DNS as
\begin{eqnarray}
\varepsilon^{\ast}(t)=E_{diss}(t)-\left(U(\{\alpha\})-U(\{\alpha_{EN}\})\right).
\end{eqnarray}
The entrance channel quantities $\{\alpha_{EN}\}$ include the proton and neutron numbers, angular momentum, quadrupole deformation parameters and orientation angles being $Z_{P}$, $N_{P}$, $Z_{T}$, $N_{T}$, $J$, $R$, $\beta_{P}$, $\beta_{T}$, $\theta_{P}$, $\theta_{T}$ for the projectile-target system. The excitation energy $E_{1}$ for fragment (Z$_{1}$,N$_{1}$) is evaluated by $E_{1}=\varepsilon^{\ast}(t=\tau_{int})A_{1}/A$. The energy dissipated into the DNS is expressed as
\begin{equation}
E_{diss}(t)=E_{c.m.}-B-\frac{\langle  J(t)\rangle(\langle J(t)\rangle+1)\hbar^{2}}{2\zeta_{rel}}-\langle  E_{rad}(J,t)\rangle.
\end{equation}
Here the $E_{c.m.}$ and $B$ are the center-of-mass energy and Coulomb barrier, respectively. The radial energy is evaluated from
\begin{equation}
\langle  E_{rad}(J,t)\rangle=E_{rad}(J,0)\exp(-t/\tau_{r})
\end{equation}
The relaxation time of the radial motion $\tau_{r}=5\times10^{-22}$ s and the radial energy at the initial state $E_{rad}(J,0)=E_{c.m.}-B-J_{i}(J_{i}+1)\hbar^{2}/(2\zeta_{rel})$. The dissipation of the relative angular momentum is described by
\begin{equation}
\langle  J(t)\rangle=J_{st}+(J_{i}-J_{st})\exp(-t/\tau_{J})
\end{equation}
The angular momentum at the sticking limit $J_{st}=J_{i}\zeta_{rel}/\zeta_{tot}$ and the relaxation time $\tau_{J}=15\times10^{-22}$ s. The $\zeta_{rel}$ and $\zeta_{tot}$ are the relative and total moments of inertia of the DNS, respectively. The initial angular momentum is set to be $J_{i}=J$ in Eq. (1). The relaxation time of radial kinetic energy and angular momentum dissipation is associated with the friction coefficients in the binary collisions.

The potential energy surface (PES) dominates the nuclear transfer and is given by
\begin{eqnarray}
U(\{\alpha\}) = && B(Z_{1},N_{1})+B(Z_{2},N_{2})-\left[B(Z,N)+V_{CN}^{rot}(J) \right]      \nonumber\\
	&& + V(\{\alpha\})
\end{eqnarray}
Here \emph{Z} and \emph{N} are the proton and neutron number of the composite system with $Z_{1}+Z_{2}=Z$ and $N_{1}+N_{2}=N$ \cite{Fe09}.  The symbol $\{\alpha\}$ denotes the quantities $Z_{1},N_{1}, Z_{2}, N_{2}; J, R; \beta_{1}, \beta_{2}, \theta_{1}, \theta_{2}$. The $B(Z_{i},N_{i}) (i=1,2)$ and $B(Z,N)$ are the negative binding energies of the fragment $(Z_{i},N_{i})$ and the compound nucleus $(Z,N)$, respectively. The $V_{CN}^{rot}$ is the rotation energy of the compound system. The $\beta_{i}$ represent the quadrupole deformations of the two fragments at ground state. The $\theta_{i}$ denote the angles between the collision orientations and the symmetry axes of deformed nuclei. The interaction potential between fragment $(Z_{1},N_{1})$ and $(Z_{2},N_{2})$ includes the nuclear, Coulomb and centrifugal parts. In the calculation, the distance $R$ between the centers of the two fragments is chosen to be the value at the touching configuration, in which the DNS is assumed to be formed. The tip-tip orientation is chosen in the calculation, which manifests the elongation shape along the collision direction and is favorable for the nucleon transfer to produce the MNT fragments.

The emission probability of preequilibrium cluster $P_{\nu}(Z_{\nu},N_{\nu},E_{k}) $ in the nucleon transfer is calculated by the uncertainty principle within the time step $t \sim  t + \triangle t$ and the kinetic energy $E_{k}$ via
\begin{equation}
P_{\nu}(Z_{\nu},N_{\nu},E_{k}) = \triangle t\Gamma_{\nu}/\hbar.
\end{equation}
Here the time step in the DNS evolution is set to be $\triangle t=0.5\times 10^{-22}$s.

The particle decay widths are evaluated with the Weisskopf evaporation theory as \cite{We37,Ch16}
\begin{eqnarray}
\Gamma_\nu(E^*,J) && = (2s_\nu + 1) \frac{m_\nu}{\pi^2 \hbar^2 \rho(E^*,J)} \int \limits ^{E^* - B_\nu - E_{rot}}_0        \nonumber   \\
&& \times \varepsilon \rho(E^*-B_\nu - E_{rot} - \varepsilon, J)\sigma_{inv}(\varepsilon) d \varepsilon .
\end{eqnarray}
Here, $s_\nu$, $m_\nu$ and $B_\nu$  are the spin, mass and binding energy of the evaporating particle, respectively. The inverse cross section is given by $\sigma_{inv}=\pi R_\nu^{2}T(\nu) $ with the radius of $R_\nu=1.21\left[(A-A_{\nu})^{1/3}+A_{\nu}^{1/3}\right]$ . The penetration probability is set to be unity for neutrons and $T(\nu) =[1 + \exp(2\pi(V_C(\nu) - \varepsilon)/\hbar\omega)]^{-1}$ for charged particles with $\hbar \omega= 5 $ and 8 MeV for hydrogen isotopes and other charged particles, respectively. Shown in Fig. 1 is a comparison of the partial decay widths of neutron, proton, deuteron, triton, $^{3}$He, $\alpha$, $^{6,7}$Li and $^{8,9}$Be from the decay of $^{221}$Ac. It can be classified three kinds of particle emission according to the magnitude, namely, neutron with the most probable emission, hydrogen isotopes and $\alpha$, other charged particles.
In the nuclear collisions, the preequilibrium clusters might be emitted from all possible DNS fragments within the dissipation of relative motion energy and angular momentum.

\begin{figure*}
\includegraphics[width=16 cm]{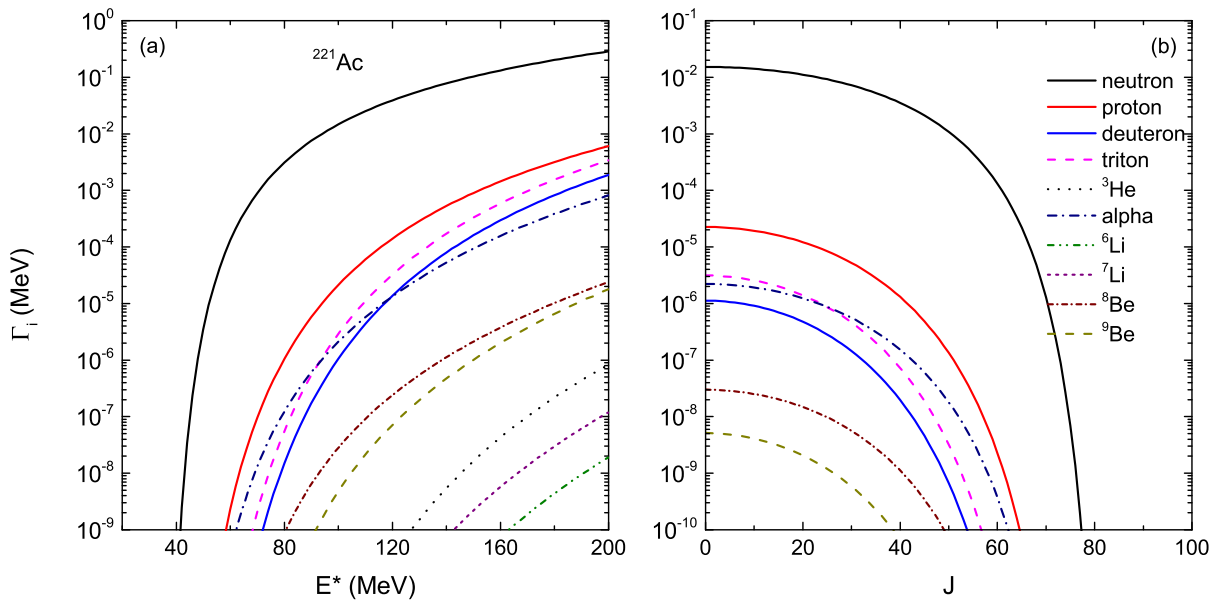}
\caption{Excitation energy and angular momentum dependence of partial decay widths of neutron, proton, deuteron, triton, $^{3}$He, $\alpha$, $^{6}$Li, $^{7}$Li, $^{8}$Be and $^{9}$Be for the decay of $^{221}$Ac. }
\end{figure*}

The level density is calculated from the Fermi-gas model \cite{Ig79} as,
\begin{eqnarray}
\rho(E^{\ast},J) && = \frac{2J+1}{24\sqrt{2}\sigma^3a^{1/4}(E^{\ast} - \delta)^{5/4}}   \nonumber \\
&& \times\exp\left[ 2\sqrt{a(E^{\ast}-\delta)} - \frac{(J+1/2)^2}{2\sigma^2}\right]
\end{eqnarray}
with $\sigma^2 = 6\bar{m}^2\sqrt{a(E^*-\delta)}/\pi^2$  and $\bar{m}\approx0.24A^{2/3}$. The pairing correction energy $\delta$ is set to be $12/\sqrt{A}, 0, -12/\sqrt{A}$  for even-even, even-odd and odd-odd nuclei, respectively. The level density parameter is related to the shell correction energy $E_{sh}(Z,N)$ and the excitation energy $E^{\ast}$ of the nucleus as
\begin{equation}
a(E^{\ast},Z,N)=\tilde{a}(A)[1+E_{sh}(Z,N)f(E^{\ast}-\Delta)/(E^{\ast}-\Delta)].
\end{equation}
Here, $\tilde{a}(A)=\alpha A + \beta  A^{2/3}b_{s}$ is the asymptotic Fermi-gas value of the level density parameter at high excitation energy. The shell damping factor is given by
\begin{equation}
f(E^{\ast})=1-\exp(-\gamma E^{\ast})
\end{equation}
with $\gamma=\tilde{a}/(\epsilon  A^{4/3})$. The parameters $\alpha$, $\beta$, $b_{s}$ and $\epsilon$ are taken to be 0.114, 0.098, 1. and 0.4, respectively \cite{Fe09}.

Once the emission probability of preequilibrium particle is determined, the kinetic energy is sampled by the Monte Carlo method within the energy range $\epsilon_{\nu}\in(0, E^* - B_\nu - E_{rot})$. The Watt spectrum is used for the neutron emission \cite{Ro92} and expressed as
\begin{equation}
\frac{dN_{n}}{d\epsilon_{n}}=C_{n}\frac{\epsilon_{n}^{1/2}}{T_{w}^{3/2}}\exp\left(-\frac{\epsilon_{n}}{T_{w}}\right)
\end{equation}
with the width $T_{w}=1.7\pm0.1$ MeV and normalization constant $C_{n}$. For the charged particles, the Boltzmann distribution is taken into account as
\begin{equation}
\frac{dN_{\nu}}{d\epsilon_{\nu}}=8\pi E_{k} \left(\frac{m}{2\pi T_{\nu}}\right)^{1/2} \exp\left(-\frac{\epsilon_{\nu}}{T_{\nu}}\right).
\end{equation}
The temperature of mother nucleus is given by $T_{\nu}=\sqrt{E^{\ast}/a}$ with the $a$ being the level density parameter.

The polar angles of preequilibrium clusters emitted from the DNS fragments are calculated by the deflection function method, which is composed of the Coulomb and nuclear deflection as \cite{Wo78,Pe22}
\begin{eqnarray}
\Theta(l_{i}) = \Theta_{C}(l_{i}) + \Theta_{N}(l_{i}).
\end{eqnarray}
The Coulomb deflection is given by the Rutherford function as
\begin{eqnarray}
\Theta(l_{i})_{C}=2\arctan \frac{Z_{1}Z_{2}e^{2}}{2E_{c.m.}b}
\end{eqnarray}
and the nuclear deflection
\begin{eqnarray}
\Theta(l_{i})_{N} = -\beta\Theta_{C}^{gr}(l_{i}) \frac{l_{i}}{l_{gr}}(\frac{\delta}{\beta})^{l_{i}/l_{gr}}.
\end{eqnarray}
Here $\Theta_{C}^{gr}(l_{i})$ is the Coulomb scattering angle at the grazing angular momentum $l_{gr}$ and $l_{gr}=0.22R_{int}[A_{red}(E_{c.m.}-V(R_{int}))]^{1/2}$. The $l_{i}$ is the incident angular momentum. The $A_{red}$ and $V(R_{int})$ are the reduced mass of projectile and target nuclei and interaction potential with $R_{int}$ being the Coulomb radius, respectively. The parameters $\delta$ and $\beta$ are parameterized by fitting the deep inelastic scattering in massive collisions as
\begin{eqnarray}
\label{eq10}
\beta = &&  75 f(\eta) + 15,   \qquad \qquad\qquad  \eta < 375      \nonumber       \\
&& 36 \exp(-2.17\times 10 ^{-3} \eta),   \qquad  \eta \geq 375
\end{eqnarray}
and
\begin{eqnarray}
\label{eq11}
\delta = && 0.07 f(\eta) + 0.11,     \qquad \qquad\qquad  \eta < 375        \nonumber   \\
&&  0.117 \exp(-1.34\times 10 ^{-4} \eta),  \qquad  \eta \geq 375
\end{eqnarray}
with
\begin{equation}
\label{eq13}
f(\eta) = [1 + \exp{\frac{\eta-235}{32}}]^{-1}.
\end{equation}
The Sommerfeld parameter $\eta = \frac{Z_1 Z_2 e^2}{\upsilon}$ and the relative velocity $\upsilon = \sqrt{\frac{2}{A_{red}}(E_c.m. - V(R_{int}))}$. For the $i-$th DNS fragment, the emission angle is determined by $\Theta_{i}(l_{i}) = \Theta(l_{i}) \xi_{i}/(\xi_{1}+\xi_{2})$ with the moment of inertia $\xi_{i}$.

\section{III. Results and discussion}

The preequilibrium clusters in the transfer reactions are associated with the nuclear structure of collision partners, i.e., the preformation factor, stiffness of nuclear surface, binding energy, coupling to the core nucleus etc, also related to the reaction dynamics, i.e., the dissipation of relative motion and coupling to the internal degrees of freedom of reaction system. The preequilibrium emission is also helpful for understanding the reaction mechanism of multinucleon transfer process, e.g., the fragment cross section, total kinetic energy configuration, angular distribution etc. Shown in Fig. 2 is the temporal evolution of the preequilibrium clusters produced in collisions of $^{48}$Ca+$^{238}$U, $^{238}$U+$^{238}$U and $^{238}$U+$^{248}$Cm at the beam energies of 8, 7 and 7.5 MeV/nucleon, respectively. The configuration of emission rate is related with the reaction system and incident energy. The reaction of $^{48}$Ca+$^{238}$U leads to the formation of compound nucleus (copernicium) and undergoes the several hundreds of 10$^{-22}$s evolution. The local excitation energy of DNS increases with the reaction time and clusters might be continuously emitted during the fusion process. The probability is small and below 0.1$\%$. The heavy systems $^{238}$U+$^{238}$U and $^{238}$U+$^{248}$Cm have the short interaction time and the cluster emission rate manifests the maximal value at the time step of 20-40$\times10^{-22}$s. The neutron production in the reactions is dominant and the emission of hydrogen isotopes is comparable with alpha in magnitude. The maximal emission rates of the preequilibrium clusters in the reactions of $^{48}$Ca+$^{238}$U and $^{238}$U+$^{238}$U are similar but different sustainable time. The preequilibrium emission is favorable with increasing the incident energy.

\begin{figure*}
\includegraphics[width=16 cm]{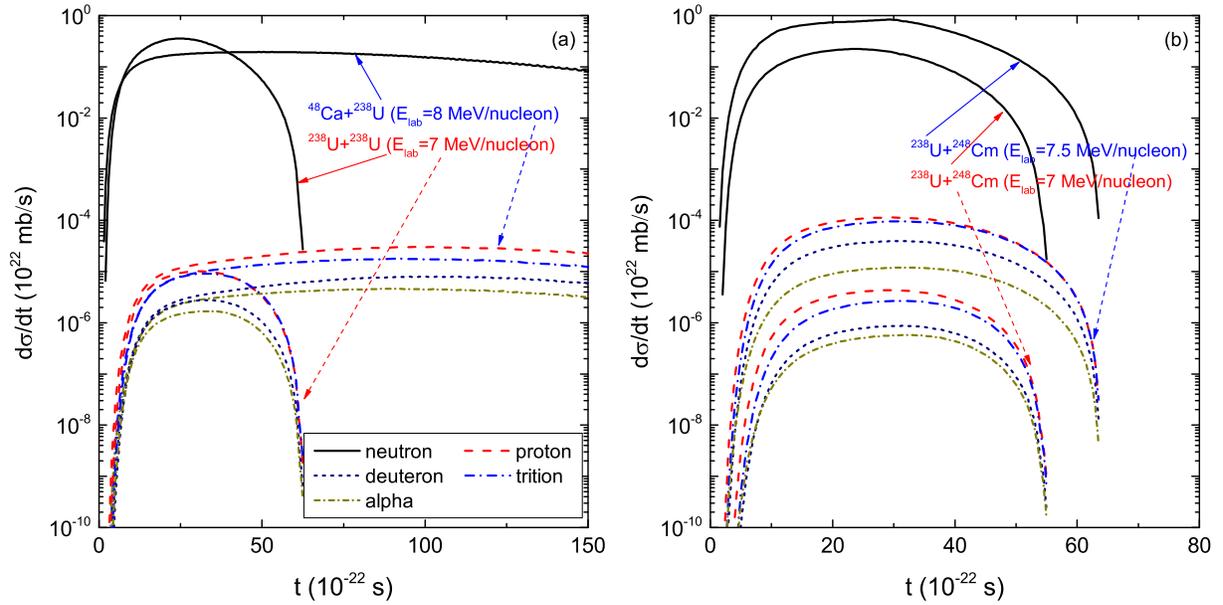}
\caption{Temporal evolution of the preequilibrium cluster emission in the reactions of $^{48}$Ca+$^{238}$U, $^{238}$U+$^{238}$U and $^{238}$U+$^{248}$Cm. }
\end{figure*}

It is well known that the emission of the MNT fragments is anisotropic and related to the reaction system and beam energy. The angular distribution of preequilibrium clusters is helpful for investigating the anisotropy of primary fragments in the MNT reactions. The sticking interaction time, moment of inertia, angular momentum, Coulomb and nuclear deflection of entrance system etc influences the emission angles of clusters corresponding to the collision orientation. We compared the angular distributions of the preequilibrium clusters produced in collisions of $^{238}$U+$^{238}$U and $^{238}$U+$^{248}$Cm at the beam energy of 7 MeV/nucleon as shown in Fig. 3.
There exists a window with 60$^{o}$-110$^{o}$ for the preequilibrium emission. The shape is very similar to the MNT fragments. Accurate estimation of emission angle is helpful for managing the detector system in experiments. In this work, we treat the preequilibrium clusters emitted from the primary fragments in the MNT reactions. Neutron, proton, deuteron, triton and alpha might be created with the same primordial nuclide.

\begin{figure*}
\includegraphics[width=16 cm]{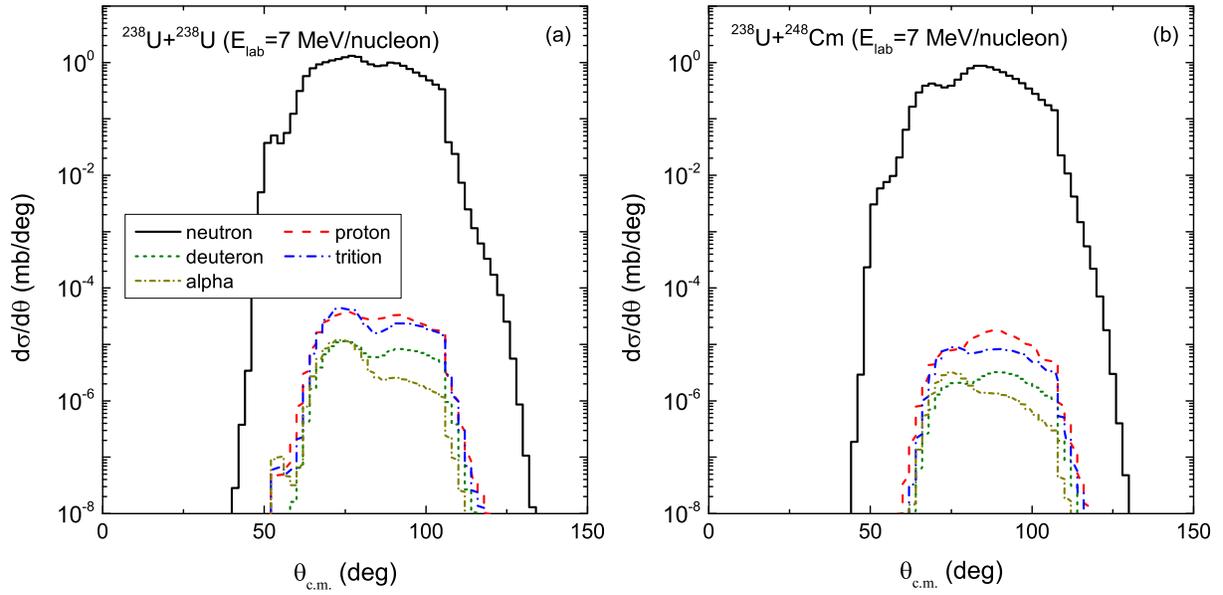}
\caption{Comparison of the angular distributions of the preequilibrium clusters produced in collisions of $^{238}$U+$^{238}$U and $^{238}$U+$^{248}$Cm at the beam energy of 7 MeV/nucleon. }
\end{figure*}

\begin{figure*}
	\includegraphics[width=16 cm]{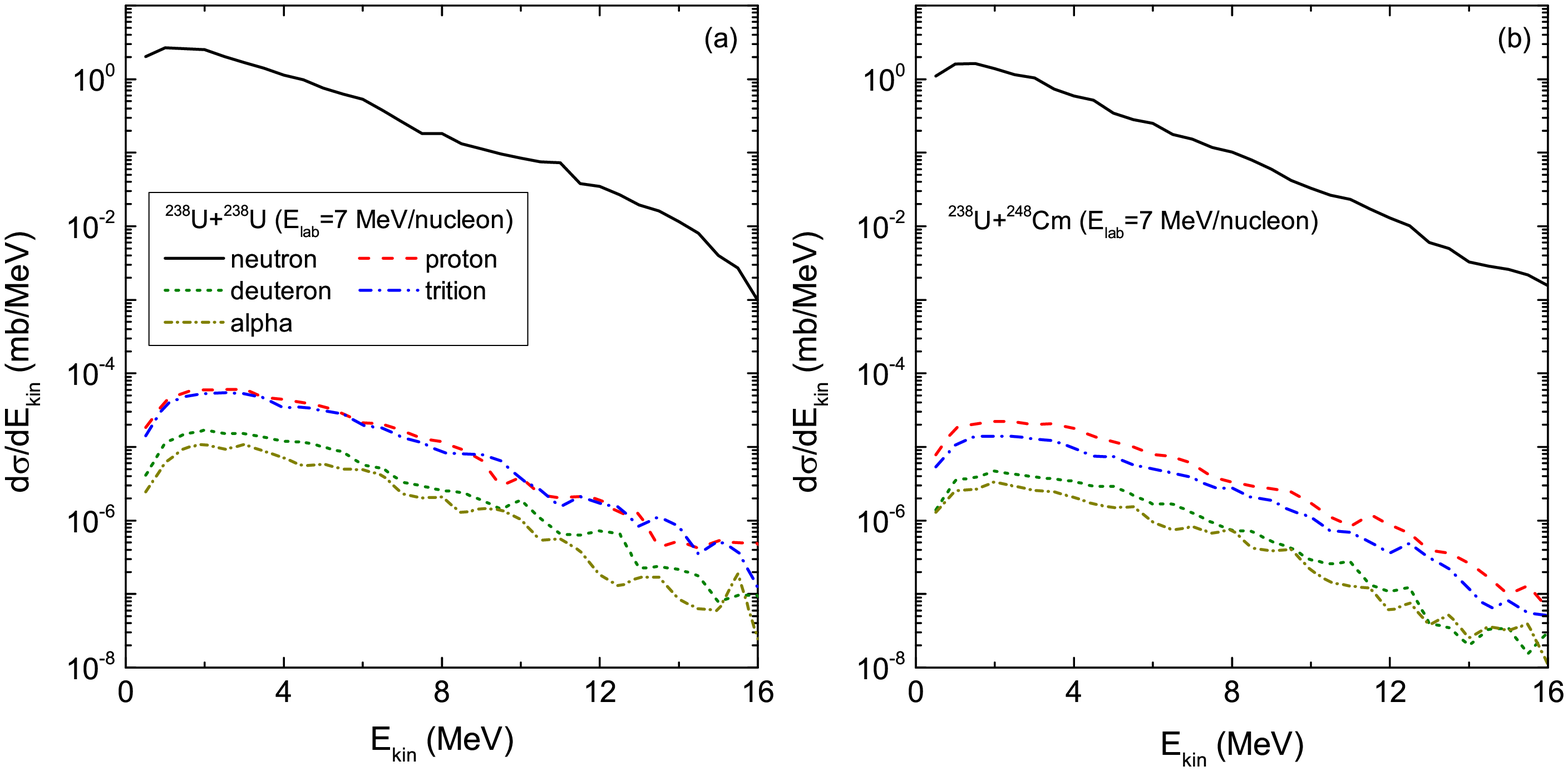}
	\caption{Kinetic energy spectra of the preequilibrium clusters produced in collisions of $^{238}$U+$^{238}$U and $^{238}$U+$^{248}$Cm. }
\end{figure*}

The kinetic energy or momentum distributions of the preequilibrium clusters in transfer reactions manifest the internal structure of cluster inside the nucleus and are also associated with the reaction dynamics. The excitation of binary nuclides, transition probability in nucleon transfer, binding energy and separation energy of cluster influence the energy spectra. Shown in Fig. 4 is a comparison of the preequilibrium neutron, proton, deuteron, triton and alpha produced in collisions of $^{238}$U+$^{238}$U and $^{238}$U+$^{248}$Cm at the incident energy of 7 MeV/nucleon. The clusters are emitted from the projectile-like or target-like fragments in the transfer reactions and manifest the Boltzmann distribution. The PES influences the local excitation energy of DNS and consequently contributes the emission probability of preequilibrium clusters. It is obvious that the neutron emission is dominant and other particles are comparable in magnitude. The distribution structure is very similar to the kinetic energy spectra in high-energy proton induced spallation reactions \cite{Ch21}. The incident energy dependence of the preequilibrium clusters is shown in Fig. 5 for the angular distributions and kinetic energy spectra in the reaction of $^{238}$U+$^{248}$Cm. The preequilibrium clusters at the energy of 7.5 MeV/nucleon are enhanced over the one-order magnitude and tend to the forward emission in comparison with the cases at 7.0 MeV/nucleon. We neglect the formation probability of cluster inside the DNS system and take the unit for all species of clusters. It has been known that the preformation of a cluster in single nucleus is described by the wave function method.

\begin{figure*}
	\includegraphics[width=16 cm]{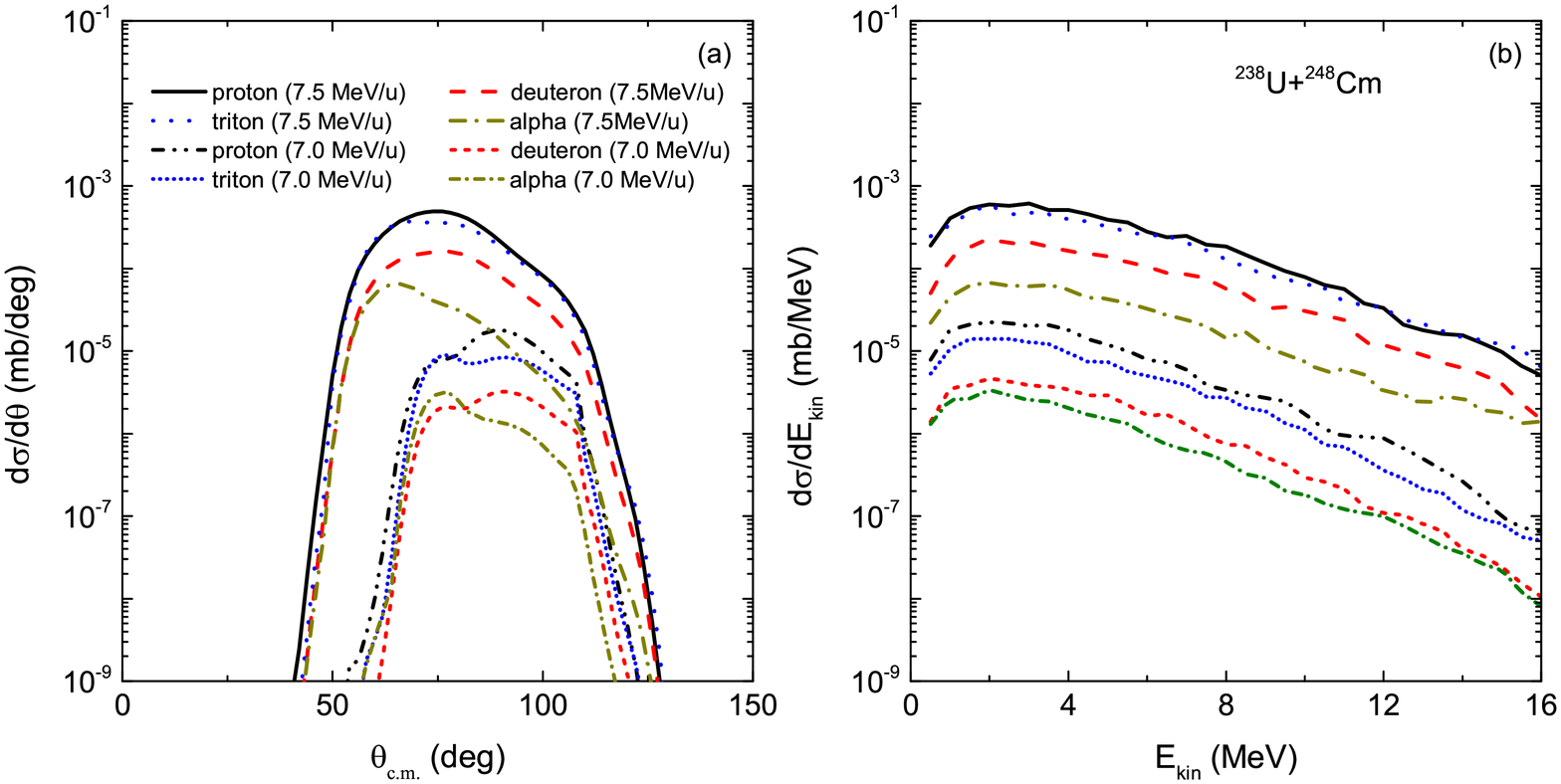}
	\caption{ (a) The angular distributions and (b) kinetic energy spectra of preequilibrium n, p, d, t and $\alpha$ in the reaction of $^{238}$U+$^{248}$Cm at the beam energies of 7 MeV/nucleon and 7.5 MeV/nucleon, respectively. }
\end{figure*}

\begin{figure*}
\includegraphics[width=16 cm]{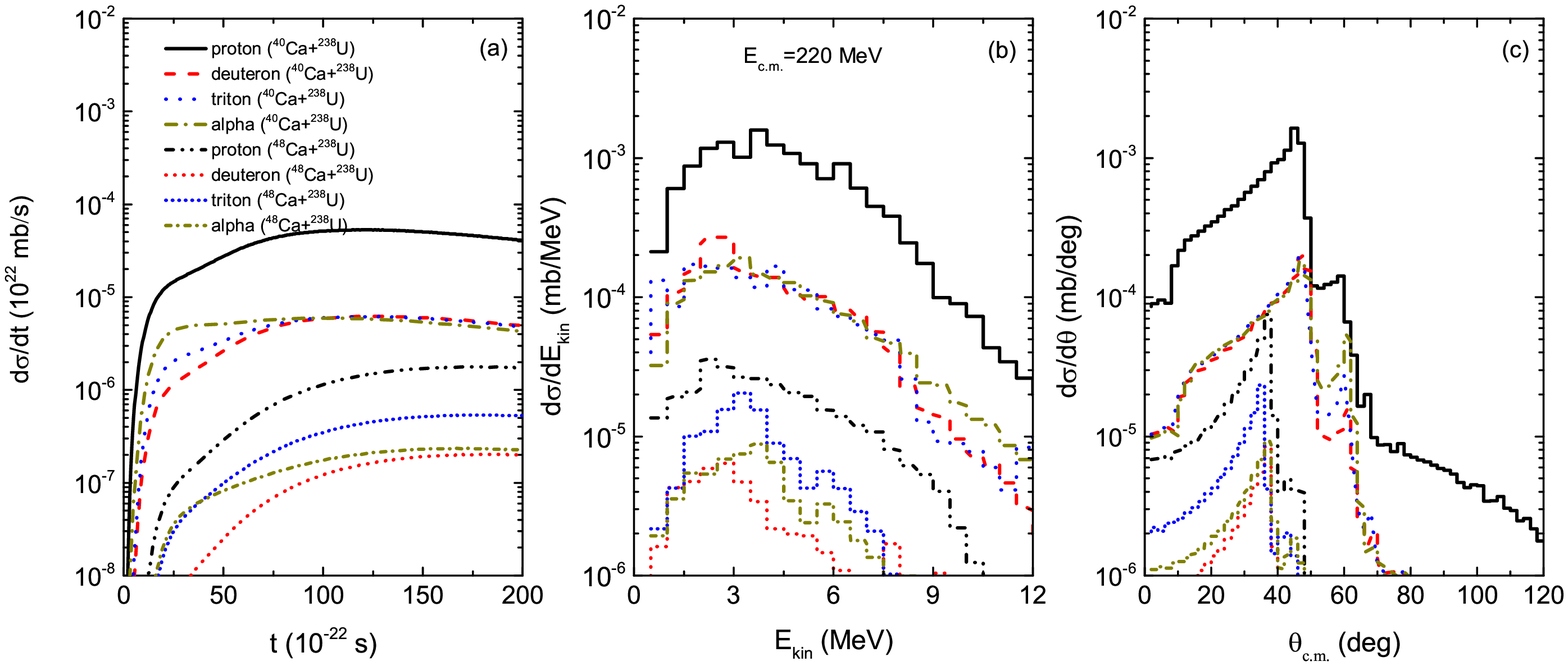}
\caption{Comparison of the production rate, kinetic energy spectra and angular distributions of preequilibrium clusters in the reactions of $^{40}$Ca+$^{238}$U and $^{48}$Ca+$^{238}$U at the center of mass energy 220 MeV. }
\end{figure*}

The cluster emission is associated with the nuclear structure and reaction dynamics. It provides the information of the single particle and multinucleon correlation of nuclear states and might be used for exploring the nuclear spectroscopics. The emission mechanism is different with the reaction system and beam energy. Shown in Fig. 6 is a comparison of the preequilibrium cluster production in the reactions of $^{40}$Ca+$^{238}$U and $^{48}$Ca+$^{238}$U at the center of mass energy 220 MeV. It is pronounced that the system $^{40}$Ca+$^{238}$U is favorable for the cluster production and has the broad energy and angular distributions.
The total cross sections of preequilibrium neutron, proton, deuteron, triton, $^{3}$He, $\alpha$, $^{7}$Li, and $^{8}$Be produced in the transfer reactions of $^{12}$C+$^{209}$Bi, $^{40,48}$Ca+$^{238}$U and $^{238}$U+$^{238}$U/$^{248}$Cm are listed in Table I. It can be classified three species according to the cross sections, the most probable emission for neutron, the medium for hydrogen isotopes and $\alpha$ with the 4-5 order lower than the neutron emission, and the lowest probability for $^{3}$He, $^{7}$Li and $^{8}$Be production. The method is also possible for the weakly bound nuclei induced reactions with the inclusion of breakup probability.

\begin{table*}
\caption{\label{tab1} Production cross sections of neutron, proton, deuteron, triton, $^{3}$He, $\alpha$, $^{7}$Li, and $^{8}$Be in the preequilibrium process of massive transfer reactions. }
\begin{ruledtabular}
\begin{tabular}{ccccccccccccc}
&system  &E$_{c.m.}$ (MeV)  &$\sigma_{n}$ (mb)    &$\sigma_{p}$ (mb)   &$\sigma_{d}$ (mb)  &$\sigma_{t}$ (mb)   &$\sigma_{^{3}He}$ (mb)   &$\sigma_{\alpha}$ (mb)    &$\sigma_{7Li}$ (mb)   &$\sigma_{8Be}$ (mb)      \\
\hline
&$^{12}$C+$^{209}$Bi   &69   &2.63    &0.26$\times10^{-3}$  &0.12$\times10^{-4}$   &0.41$\times10^{-4}$   &0.62$\times10^{-11}$   &0.22$\times10^{-3}$    &0.19$\times10^{-12}$   &0.42$\times10^{-12}$        \\
&$^{40}$Ca+$^{238}$U  &220  &24.65       &0.15$\times10^{-1}$   &0.17$\times10^{-2}$   &0.18$\times10^{-2}$     &0.62$\times10^{-8}$   &0.20$\times10^{-2}$   &0.16$\times10^{-10}$   &0.53$\times10^{-11}$    \\
&$^{48}$Ca+$^{238}$U  &180   &0.11$\times10^{-1}$   &0.84$\times10^{-12}$   &0.21$\times10^{-14}$   &0.24$\times10^{-13}$   &$<10^{-16}$       &0.77$\times10^{-13}$    &$<10^{-16}$     &$<10^{-16}$     \\
&$^{48}$Ca+$^{238}$U  &200   &1.63    &0.42$\times10^{-5}$   &0.18$\times10^{-6}$   &0.66$\times10^{-6}$ &$<10^{-16}$      &0.54$\times10^{-6}$       &$<10^{-16}$     &$<10^{-16}$     \\
&$^{48}$Ca+$^{238}$U  &220   &23.16      &0.40$\times10^{-3}$     &0.44$\times10^{-4}$   &0.12$\times10^{-3}$   &0.23$\times10^{-11}$   &0.60$\times10^{-4}$   &0.74$\times10^{-14}$   &0.53$\times10^{-15}$    \\
&$^{48}$Ca+$^{238}$U  &240   &96.11      &0.61$\times10^{-2}$     &0.11$\times10^{-2}$   &0.28$\times10^{-2}$   &0.61$\times10^{-9}$   &0.94$\times10^{-3}$   &0.48$\times10^{-11}$   &0.34$\times10^{-12}$   \\
&$^{238}$U+$^{238}$U  &833   &20.59       &0.61$\times10^{-3}$   &0.17$\times10^{-3}$   &0.55$\times10^{-3}$   &0.49$\times10^{-10}$   &0.11$\times10^{-3}$   &0.15$\times10^{-11}$   &0.11$\times10^{-12}$   \\
&$^{238}$U+$^{248}$Cm  &850  &11.53      &0.23$\times10^{-3}$     &0.46$\times10^{-4}$   &0.14$\times10^{-3}$   &0.63$\times10^{-11}$   &0.31$\times10^{-4}$   &0.20$\times10^{-12}$   &0.16$\times10^{-13}$     \\
&$^{238}$U+$^{248}$Cm  &911  &56.27       &0.71$\times10^{-2}$   &0.24$\times10^{-2}$   &0.60$\times10^{-2}$    &0.41$\times10^{-8}$   &0.77$\times10^{-3}$   &0.10$\times10^{-9}$   &0.56$\times10^{-11}$   \\

\end{tabular}
\end{ruledtabular}
\end{table*}

\section{IV. Conclusions}

In summary, the emission mechanism of preequilibrium clusters in collisions of $^{12}$C+$^{209}$Bi, $^{40,48}$Ca + $^{238}$U, $^{238}$U+$^{238}$U and $^{238}$U+$^{248}$Cm near Coulomb barrier energies has been systematically investigated within the DNS model. The preequilibrium clusters are considered to be emitted from the decay of the primordial DNS fragments in the nucleon transfer process. The production rate is associated with the reaction system and beam energy. The preequilibrium emission takes place until the formation of compound nucleus. The kinetic spectra manifest the Boltzmann shape. The angular distribution is similar to the transfer fragments, i.e., in the range $70^{o}-110^{o}$ for the reactions of $^{238}$U+$^{238}$U/$^{248}$Cm, forward emission $35^{o}-60^{o}$ for the light systems of $^{40,48}$Ca + $^{238}$U. The production cross sections of preequilibrium clusters strongly depend on the separation energy and Coulomb barrier from the primordial nuclides. The neutrons are emitted and also take away the local excitation energy of the DNS system. The emission rate of alpha and hydrogen isotopes is comparable in the magnitude. The production of heavier clusters, such as lithium, beryllium isotopes etc, are associated with the reaction system. The method is also possible for describing the weakly bound nuclei induced reactions.

\section{Acknowledgements}
This work was supported by the National Natural Science Foundation of China (Projects No. 12175072 and No. 11722546) and the Talent Program of South China University of Technology (Projects No. 20210115).

\end{document}